\documentclass[graybox]{svmult}

\usepackage{mathptmx}       
\usepackage{helvet}         
\usepackage{courier}        
\usepackage{type1cm}        
\usepackage{algorithmic}

\usepackage{makeidx}         
\usepackage{graphicx}        
\usepackage{multicol}        
\usepackage[bottom]{footmisc}

\usepackage{url}
\usepackage{tabularx}
\usepackage{amsmath}
\usepackage{subfigure}
\usepackage{verbatim}
\usepackage{multirow}

\makeindex

\begin{document}
\title*{Advancing Intra-operative Precision: Dynamic Data-Driven Non-Rigid Registration for Enhanced Brain Tumor Resection in Image-Guided Neurosurgery}

\titlerunning{Non-Rigid Registration for Brain Tumor Resection in Image-Guided Neurosurgery}
\authorrunning{Chrisochoides et al.}

\author{Nikos Chrisochoides, Andriy Fedorov, Fotis Drakopoulos, Andriy Kot, Yixun Liu, Panos Foteinos, Angelos Angelopoulos, Olivier Clatz, Nicholas Ayache, Peter M. Black,
Alex J. Golby, Ron Kikinis}
\institute{
Nikos Chrisochoides, Andriy Fedorov, Fotis Drakopoulos, Panos Foteinos, Angelos Angelopoulos, Andriy Kot, Yixun Liu \at Center for Real-Time
Computing, Old Dominion University, Norfolk,  VA USA
\email{nikos@cs.odu.edu}
\and
Andriy Fedorov, Ron Kikinis, Alexandra J. Golby \at Surgical Planning Laboratory,
Department of Radiology,
Brigham and Women's Hospital, Boston, MA USA
\and
Olivier Clatz, Nicholas Ayache \at INRIA Sophia Antipolis, France
\and
Alexandra J. Golby, Peter M. Black \at Department of Neurosurgery, Brigham and
Women's Hospital, Boston, MA USA
}

\maketitle

\abstract*{During neurosurgery, medical images of the brain are used to locate tumors and critical structures, but brain tissue shifts make pre-operative images unreliable for accurate removal of tumors. Intra-operative imaging can track these deformations but is not a substitute for pre-operative data. To address this, we use Dynamic Data-Driven Non-Rigid Registration (NRR), a complex and time-consuming image processing operation that adjusts the pre-operative image data to account for intra-operative brain shift. Our review explores a specific NRR method for registering brain MRI during image-guided neurosurgery and examines various strategies for improving the accuracy and speed of the NRR method.  We demonstrate that our implementation enables NRR results to be delivered within clinical time constraints while leveraging Distributed Computing and Machine Learning to enhance registration accuracy by identifying optimal parameters for the NRR method. Additionally, we highlight challenges associated with its use in the operating room.}

\abstract{During neurosurgery, medical images of the brain are used to locate tumors and critical structures, but brain tissue shifts make pre-operative images unreliable for accurate removal of tumors. Intra-operative imaging can track these deformations but is not a substitute for pre-operative data. To address this, we use Dynamic Data-Driven Non-Rigid Registration (NRR), a complex and time-consuming image processing operation that adjusts the pre-operative image data to account for intra-operative brain shift. Our review explores a specific NRR method for registering brain MRI during image-guided neurosurgery and examines various strategies for improving the accuracy and speed of the NRR method.  We demonstrate that our implementation enables NRR results to be delivered within clinical time constraints while leveraging Distributed Computing and Machine Learning to enhance registration accuracy by identifying optimal parameters for the NRR method. Additionally, we highlight challenges associated with its use in the operating room.}


\section{Introduction}

Cancer is one of the leading causes of death in the USA and worldwide.  Among the different types of cancer, brain cancer was estimated to
claim over 50 thousand new victims in 2008~\cite{abta_facts}, when we first consider summarizing the preliminary results of our approach. However, brain cancer continues to be a significant healthcare problem. The number of Americans living with brain tumors exceeds 700,000, surpassing the number of COVID-19 deaths in mid-summer 2021.

Neurosurgical resection is one of brain tumor patients' most common and effective treatment
options. The resection must remove
as much as possible of the tumor tissue while maximally preserving the
vital structures of the healthy brain. Maximal tumor excision increases time
to progression, and reduces symptoms and seizures. 
In this Chapter, we explore how the concept of Dynamic Data Driven
Applications Systems (DDDAS)~\cite{dddas, dddas_book}, together with the advances in
medical image acquisition,  distributed computing, and Machine Learning can assist in enabling
image guidance during neurosurgery and potentially can improve the accuracy
of the procedure, allowing more complete tumor resections without additional
morbidity. We focus more on the implementation aspects, while in \cite{digital-health-23}, we focus more on the approach's mathematical modeling and computational aspects. 

\begin{figure}[h]
\centering
\includegraphics[width=.4\linewidth]{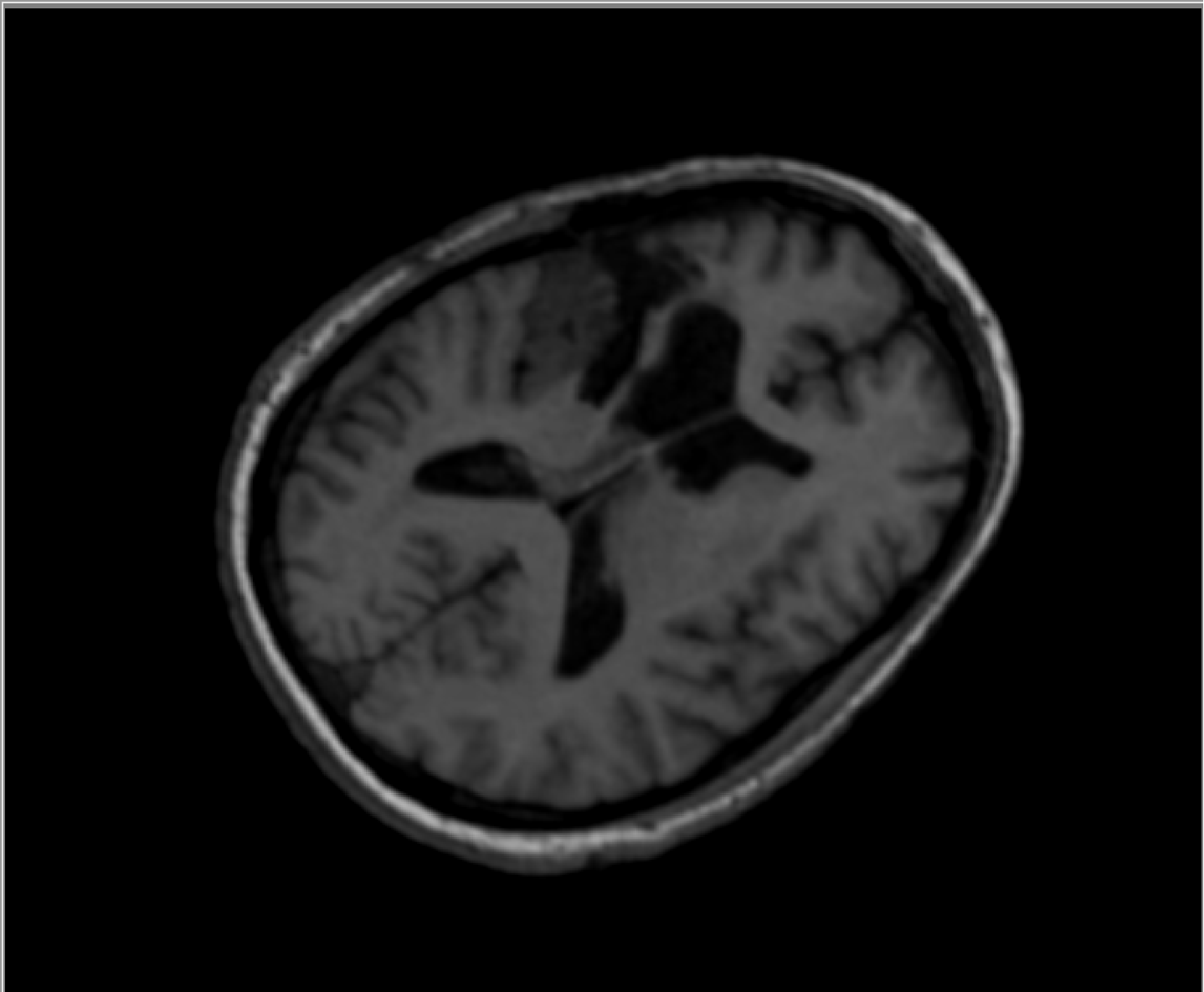}
\includegraphics[width=.4\linewidth]{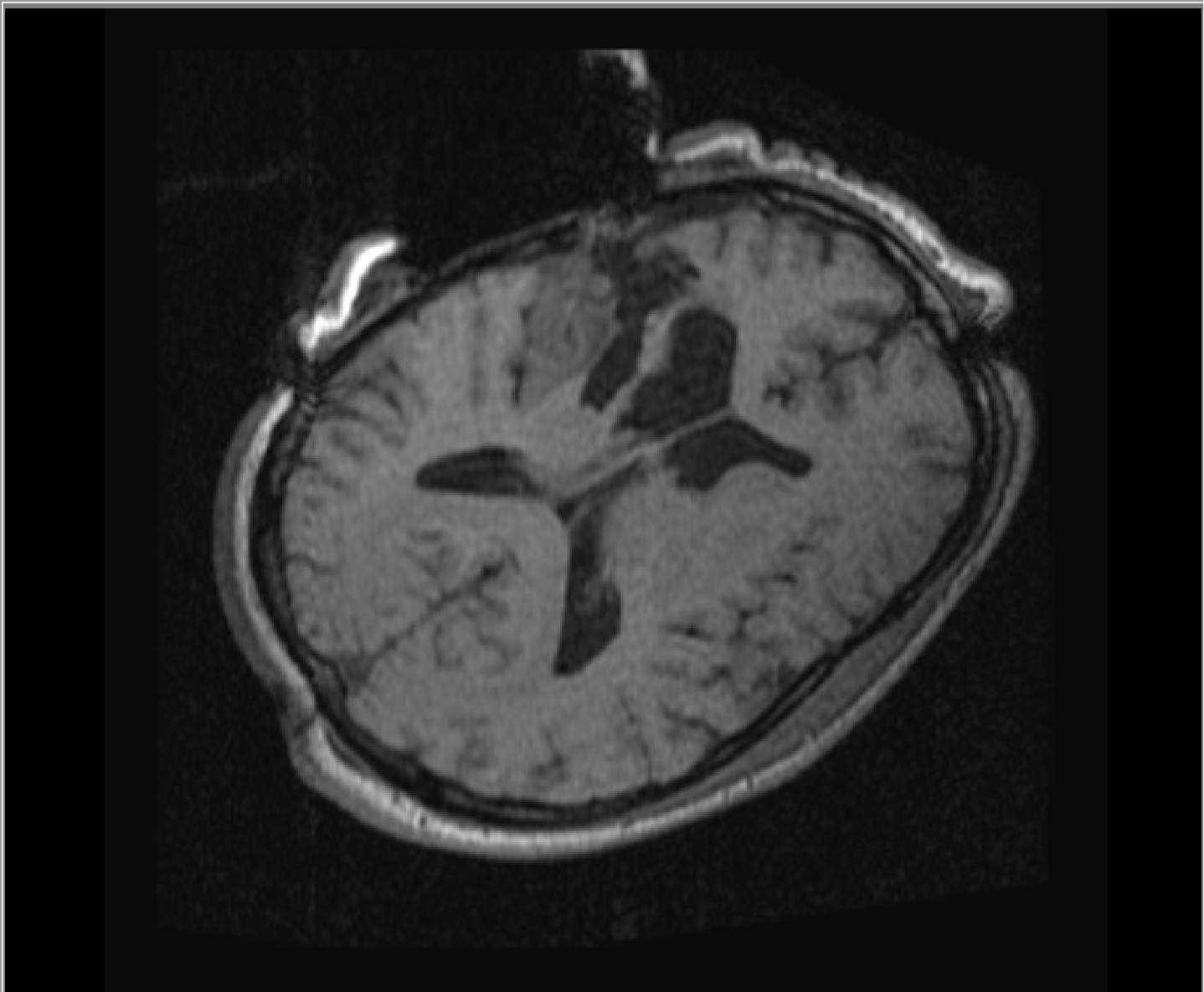}
\caption{Intra-operative brain deformation. Left: pre-operative, higher
quality image, showing the location of brain tumor. Right: intra-operative
image showing brain shift~\cite{ignsPublicCases}.\label{fig:brainshift}}
\end{figure}

There are many challenges in accomplishing the objectives of
neurosurgery. In this Chapter, we focus on one of them: the
exact locations of the brain areas responsible for critical brain
function, e.g., the motor cortex, are patient-specific and cannot be
identified with the naked eye. This is where medical imaging, distributed computing,  and machine learning become essential.

Magnetic Resonance Imaging (MRI) is indispensable in demonstrating brain
pathologies. Although not distinguishable from the naked eye, neoplastic
tissues can be differentiated from brain tissue based on changes in MR signal
and corresponding image intensities. MRI has also been shown to be useful in
constructing functional mapping of the brain using functional MRI
(fMRI)~\cite{golby02}. Both the structural and functional imaging data are
used for the purposes of improving the precision of the resection.

Image registration in general, is concerned with spatial alignment of
corresponding features in two or more images. During image registration, a
spatial transformation is applied to one image, which is called
{\emph{floating}}, so that it is brought into alignment with the
{\emph{target}}, or {\emph{reference}} position of the object (brain). During rigid
image registration, the floating image corresponds to the pre-operative image,
which is aligned with the patient's position using translations and
rotations (rigid transformations). 

During the course of surgery, opening of the skull and dura causes changes
in pressure inside the Intra-Cranial Cavity (ICC). 

Because of this and other factors, such as drainage of cerebrospinal fluid,
induced changes in brain tumors, and the effect of gravity, the brain changes its shape, introducing
discrepancies in relation to the pre-operative configuration.
{\emph{Non-rigid}} image registration uses a spatially varying transformation to
account for this deformation. 
In general, image registration algorithms are based on optimizing certain
similarity criteria between the fixed and floating image under varying
parameters of spatial transformation. The complexity of this optimization
depends on the number of parameters that describe the transformation. Both
rigid and non-rigid registration are open research areas in medical image
processing. However, non-rigid registration is a conceptually more difficult
problem, which usually requires significant computing resources and time.

Non-rigid registration recovers the deformation of the brain based on 
intra-operatively acquired imaging data.  Advances in medical image
acquisition have made it possible to acquire high-resolution images, in particular
MRI during the surgery. Intra-operative MRI (iMRI) cannot substitute
pre-operative images because of its limited resolution and the high processing
time required to obtain functional data.  However, iMRI can be used to guide
the registration of the pre-operative data.

There are three main requirements to non-rigid registration
(NRR)~\cite{clatz05}. First, NRR should deliver accurate results. Second, the
result should be consistently accurate, independent of the specific registered images, and not be sensitive to small variations in the
parameter selection. One important requirement for Image-Guided Neurosurgery  (IGNS) is that registration must be completed within the neurosurgical workflow's time constraints, typically between 4 to 5 minutes.

The prospective application of NRR is a dynamic process. iMRI is obtained
periodically as requested by the surgeon. Immediately following iMRI, NRR should
be used to estimate the deformation of the brain and update the pre-operative
images.  Usually, hospitals do not have locally available large-scale
computational facilities. In this Chapter, we describe an infrastructure that
enables the computation of non-rigid registration using remotely located high-performance computing resources guided by intra-operative image updates.

In the context of the application, we define the {\emph{response time}} as the
time between the acquisition of the intra-operative scan of the deformed
tissue and the final visualization of the registered preoperative data
on the console in the operating room. These steps are performed intra-operatively
form the DDDAS steered by the periodic
acquisitions of the iMRI data.  Our broad objective is to minimize the
perceived (end-to-end) response time of the DDDAS component. 

To our knowledge, none of the systems developed (2005  timeframe) were used
prospectively during image-guided neurosurgeries.  Our approach to developing 
such a dynamic data-driven NRR system for IGNS was to
adopt an existing NRR method of established accuracy and parallelize the
most time-consuming components of this method and develop an end-to-end
system to facilitate image guidance during neurosurgery.

\section{Related Work}

The research in NRR for IGNS can be separated into the development of the core
registration methods and the design of end-to-end systems capable of supporting NRR computation and delivering the results intra-operatively.  The choice
of the NRR method depends mostly on the intra-operative image modality that
captures brain deformation~\cite{RefWorks:1056}. However, the core computation
components of NRR are very similar for different intraoperative imaging
modalities.

Registration algorithms are based on optimizing certain similarity measures
between the intensities of the reference and floating images. In non-rigid
registration, the number of parameters (degrees of freedom) that are being
optimized is exceedingly large compared to rigid registration. This
contributes significantly to the costs of computing the similarity metric and
to the evaluation of gradients required during optimization. However,
optimizing the similarity measure alone can lead to unrealistic solutions
since non-rigid registration is an ill-posed problem. Therefore,
NRR usually includes some form of solution regularization.  Biomechanical modeling of tissue
deformation is one such regularization approach.  Deformation of tissue is usually
modeled using the Finite Element Method (FEM)~\cite{RefWorks:808}, which requires solving
a system of equations. The size of this system is proportional to the
resolution of the brain biomechanical model.

Timely completion of the core NRR computations is the key component
for efficient end-to-end registration systems. Several strategies have
been proposed to parallelize the time-consuming steps in medical image
processing.  Christensen and collaborators were some of the first to discuss
using parallel computing resources for solving time-consuming problems
related to brain MRI processing on a massively parallel SIMD
architecture~\cite{RefWorks:1031}.  

Warfield et al.~\cite{RefWorks:1029} presented some of the first results in
intra-operative processing (segmentation) of iMRI.  The authors demonstrate
linear speedup of segmentation on a 20-processor workstation, which allows 
the processing of a  typical dataset in about 20 seconds.  The developed method
was applied and evaluated prospectively during neurosurgeries and
liver cryo-ablation procedures~\cite{RefWorks:1028}. The same group later
developed a high-performance method for intra-operative non-rigid
registration, which uses linear biomechanical model~\cite{RefWorks:1027}
solved in parallel.  Although the authors report clinically acceptable timing
results delivered by their implementation, the evaluation was restricted to
off-line experimental studies.

Computation of the NRR result within the time constraints of neurosurgery is
an essential requirement. To facilitate this task,
support of the computation on the remote resources may be
required. The community has recognized these issues, and several
solutions have been proposed. Stefanescu et al.~\cite{RefWorks:920} describe an NRR
implementation that is exposed as a web service. Ino et al. developed an
end-to-end system for rigid registration computation on a remote
cluster~\cite{RefWorks:927}. Lippman and Kruggel use a customized grid infrastructure to
design an NRR system for IGNS~\cite{lippman05}. 

\begin{figure}[b]
\includegraphics[width=1.0\linewidth]{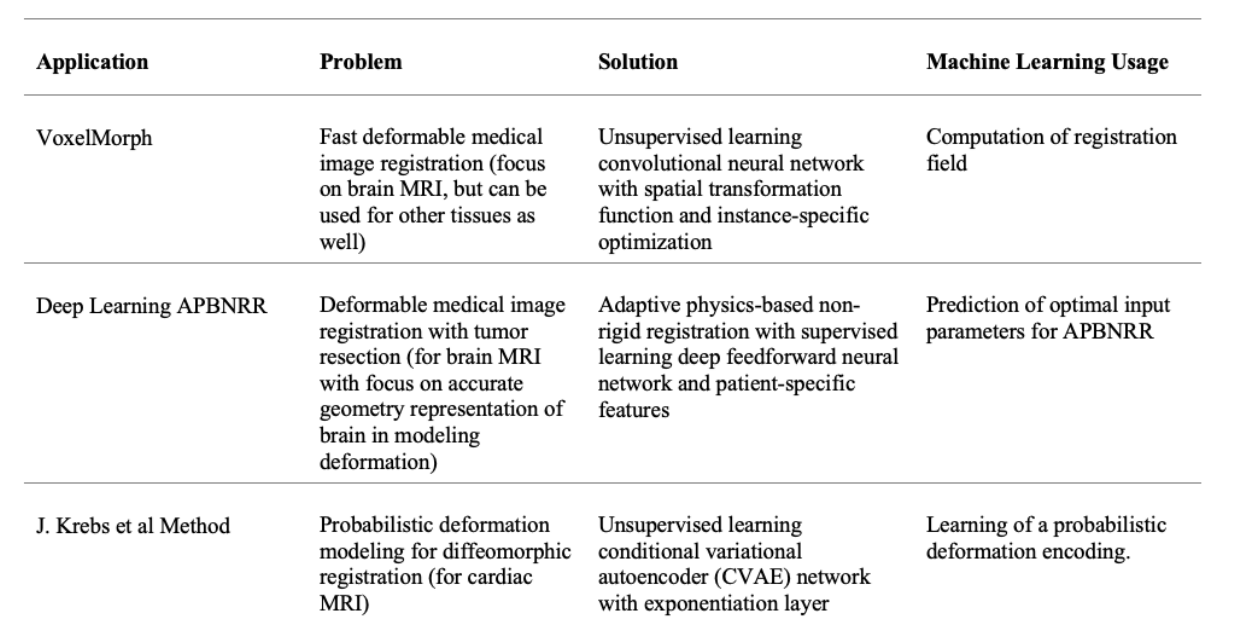}
\caption{Compares three machine learning approaches in medical image registration for major applications: two state-of-the-art and APBNRR deep learning; the columns from left to right correspond to the application, problem, solutions, and usage. \label{fig: APBNRR_Learning}}
\end{figure}

Medical image computing is one area where machine learning (ML) is applied to solve problems \cite{8241753}. Specifically, for medical image registration, many new methods currently utilize convolutional neural networks, which operate directly on the preoperative and intraoperative MRI images to produce a registered image. A notable example is MIT’s VoxelMorph model \cite{VoxelMorph}. VoxelMorph utilizes a CNN, which computes a registration field and warps the preoperative image with the registration field using a spatial transformation function.	 In VoxelMorph, ML is used directly for image registration, while in our case ML is used for optimal values of the parameters in NRR.  This Chapter will look at two major applications, VoxelMorph\cite{VoxelMorph}   and the method proposed by J. Krebs et al. \cite{krebs2018unsupervised}. 

VoxelMorph \cite{VoxelMorph} addresses the problem of fast deformable medical image registration with a focus on brain MRI, but it can also be used for other tissues.  VoxelMorph uses a solution formulated by an unsupervised learning convolutional neural network for computing a registration field and a spatial transformation function for warping the preoperative image. The application can also use instance-specific optimization by fine-tuning the network parameters for each MR image. As noted in  \cite{VoxelMorph} it runs in 0.45 seconds on a top-tier GPU and 57 seconds on a CPU, with an average DICE ~\cite{Eelbode_2020}  score of 75.3\%. Another approach is proposed by J. Krebs et al. \cite{krebs2018unsupervised} aims to address the issue of probabilistic deformation modeling for diffeomorphic registration in cardiac MRI. Their solution employs an unsupervised learning conditional variational autoencoder (CVAE) network, which utilizes an exponentiation layer to create diffeomorphic transformations. The average execution time is 0.32 seconds on a top-tier GPU with an average DICE score of 79.9\% and a mean Hausdorff distance of 7.9 mm. A comparison of the problems, the proposed solutions, and the usage of machine learning in NRR and the two applications listed above are presented in the  Table of Figure \ref{fig: APBNRR_Learning}  from \cite{Angelopoulos19D}.

\section{Physics-Based Non-Rigid Registration}

The core registration method of our dynamic infrastructure was originally
developed by Clatz et al. in~\cite{clatz05}. This Physics-Based Non-Rigid Registration (PBNRR) approach is specifically
designed for registering high-resolution pre-operative data with iMRI.  The
NRR computation consists of preoperative and intra-operative components.
Intraoperative processing starts with the acquisition of the first iMRI.
However, the {\emph{time-critical}} part of the intra-operative computation is
initiated when a scan showing the shift of the brain is available. The basic idea
of the registration method is to estimate the {\emph{sparse deformation field}}
that matches similar locations in the image and then uses a biomechanical model
of the brain, deformation to discard unrealistic displacements and derive
{\emph{dense deformation field}} that defines transformation for each point in
the image space. 

Sparse displacement vectors are obtained at the selected points in the image,
where the variability in the intensities in the surrounding region exceeds
some threshold. Such {\emph{registration}} points can be identified prior to
the time-critical part of the computation in the floating (pre-operative)
image. Once the reference (intra-operative) scan is available, the deformation
vector is estimated at each of the selected points by means of block matching.
Fixed-size rectangular regions (blocks) centered at the registration points
are identified in the floating image.  Given such a block, we next select a
search region (window) in the reference image. The block's displacement
that maximizes the intensity-based similarity metric between the image intensities
in the block and the overlapping portion of the window corresponds to the
vector value of the sparse deformation field at the registration point. The
normalized cross-correlation (NCC) similarity metric is evaluated as follows:
$$NCC=\frac{\sum_{i\in{B}}(B_T(i)-\bar{B}_T)(B_F(i)-\bar{B}_F)}{\sqrt{(B_T(i)-\bar{B}_T)^2(B_F-\bar{B}_F)^2}}.$$
$\bar{B}_T$ and $\bar{B}_F$ correspond to the average intensity values within
the block in the reference and floating image, respectively. We note the high
computational complexity of the block-matching procedure.  Considering the
sizes of three-dimensional blocks and windows are defined in pixels as
$B=\{B_x,B_y,B_z\}$ and $W=\{W_x,W_y,W_z\}$, the bound on the number of
operations is $O(B_xB_yB_z\times W_xW_yW_z)$ for one registration point.

Estimation of brain deformation is based on the finite element
analysis (FEA) using a linear elastic model of brain deformation. The finite element mesh of the
intra-cranial volume is constructed from the segmented ICC volume following
the methods we evaluated in a separate studies~\cite{RefWorks:1002} and \cite{Drakopoulos17A}. We then iteratively
seek such a position of the mesh vertices $\mathbf{U}$ that balances the
mechanical forces of the modeled tissue that resist deformation, with the
external forces that correspond to the displacements $\mathbf{D}$ estimated
by block matching:
$$\mathbf{F}_i\Leftarrow \mathbf{KU}_i,\ \mathbf{U}_{i+1}\Leftarrow [\mathbf{K}+\mathbf{H}^T\mathbf{SH}]^{-1}[\mathbf{H}^T\mathbf{SD}+\mathbf{F}_i].$$
Here, $\mathbf{K}$ is the mechanical stiffness matrix~\cite{delingette_book}, $\mathbf{H}$ is
the interpolation matrix from the mesh vertices to the block matching
displacements, $\mathbf{S}$ is the matrix that captures the confidence in the
block matching results. $\mathbf{F}$ is the force that is increasing between
iterations to slowly cancel the influence of the mechanical forces.

Both block matching and iterative estimation of displacements are time
critical and should be performed while the surgeons are waiting. Block
matching contributes most to the computation costs because of the
exhaustive search for the optimum block position. Iterative estimation
of mesh vertex displacements based on a biomechanical model requires a solution of
a system of linear equations during each iteration. However, the size of that
system is constrained by the number of mesh vertices, which cannot be
arbitrarily large due to inherent properties of the NRR
algorithm~\cite{RefWorks:1002}.

\subsection{\bf Adaptive Non-Rigid Registration (APBNRR)} 
\begin{figure}[tb]
\centering
\includegraphics[width=5in,height=3in]{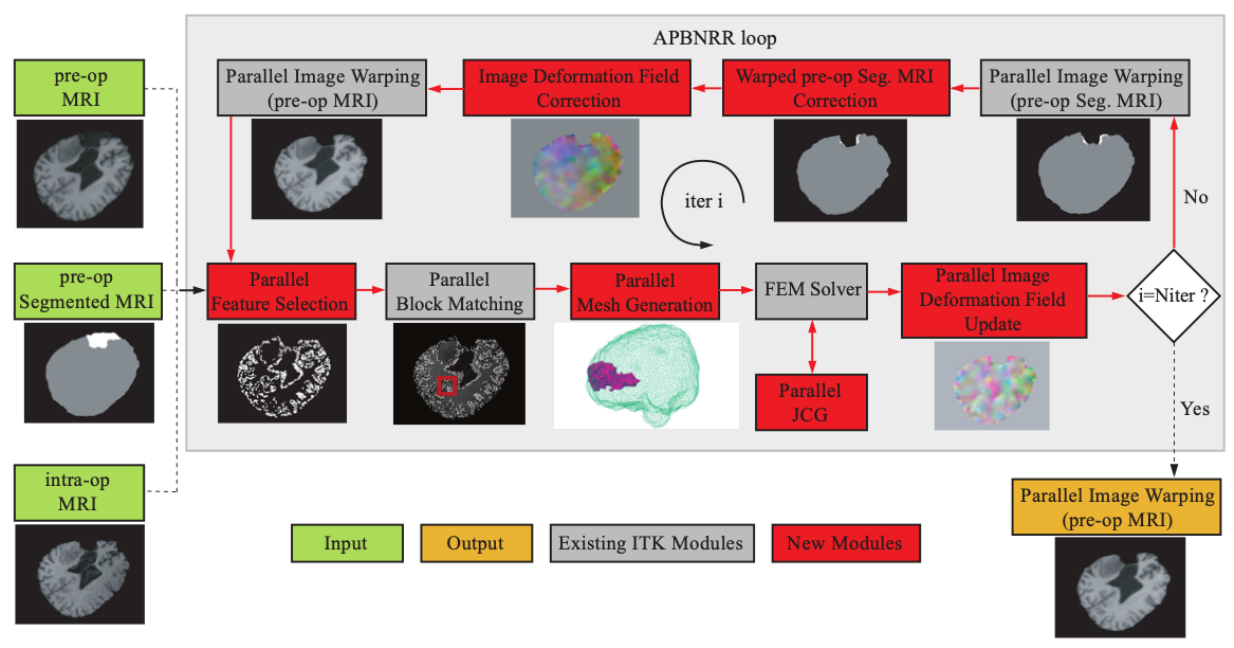}
\caption{\label{fig:APBNRR_Frame} The APBNRR framework. The green, red, and gray boxes represent the input, the additional parallel modules to manage real-time adaptivity (in red), and the existing ITK modules \cite{Liu14A}. The red arrows show the execution order of the modules. Orange represents the output warped pre-operative MRI.}
\end{figure}

However, complete resection of large brain tumors leads to large brain shifts. This creates a large cavity of elements in the tessellated brain image model which compromises the accuracy of the biomechanical model defined on pre-operative MRI. This section summarizes the extensions of the PBNRR by: (i) removing additional outliers due to tissue resection using an Adaptive Non-Rigid Registration (APBNRR) method and gradually adjusting the mesh for the FEM model to an incrementally warped segmented intraoperative  MRI (iMRI), see Figure \ref{fig:APBNRR_Frame} and for a more detailed description see \cite{Drakopoulos14A},  \cite{Drakopoulos15A} and \cite{Drakopoulos17A}.

\begin{figure}[tb]
\centering
\includegraphics[width=5in,height=5in]{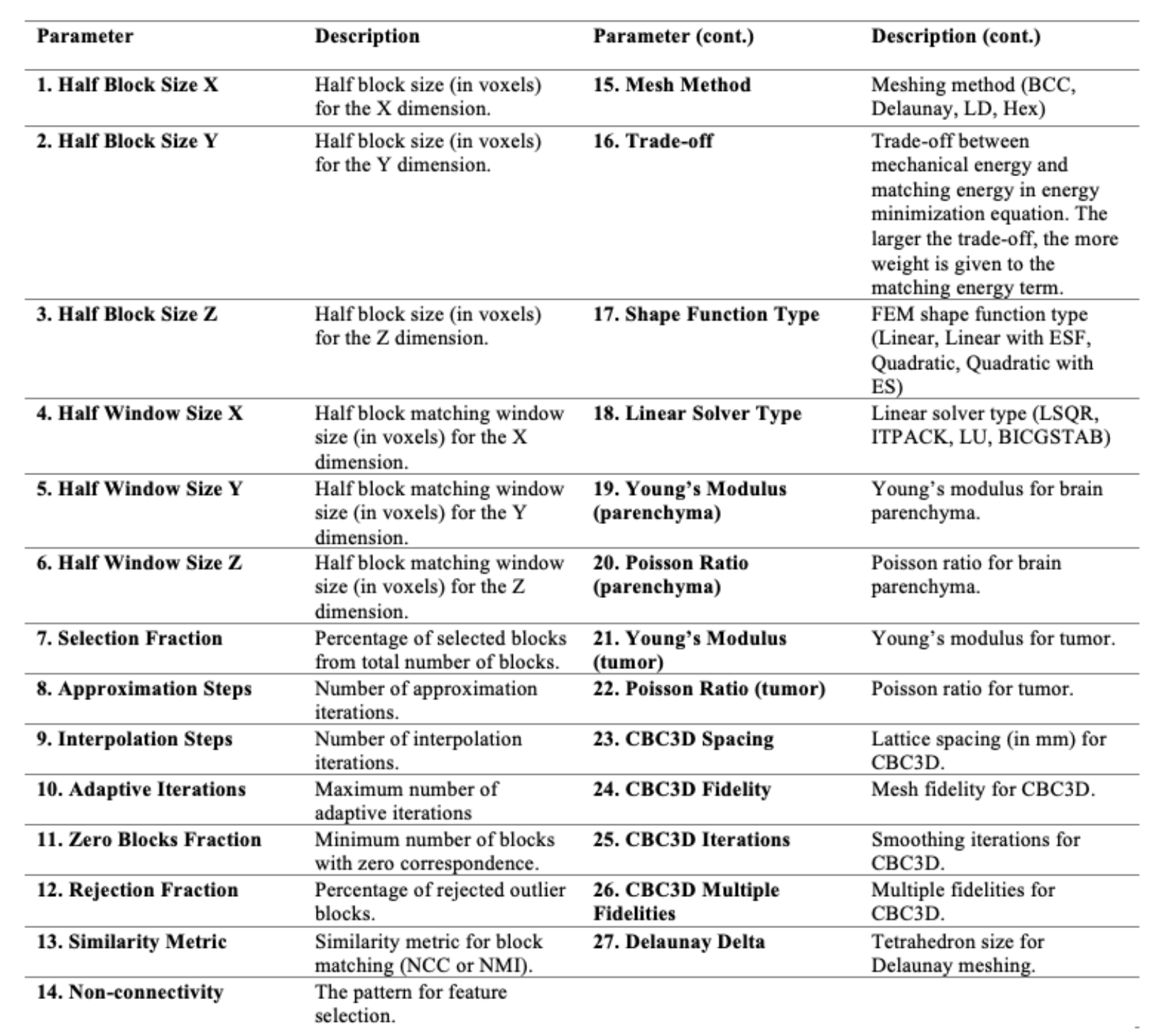}
\caption{\label{fig:params} Shows the tunable parameters APBNRR utilizes. Parameters 1-12 are utilized in the deep learning model, while the rest are fixed. I/O parameters are not included in this table. CBC3D refers to the latest image-to-mesh conversion presented in \cite{Drakopoulos15T}.}
\end{figure}

APBNRR takes as input a preoperative segmented MRI (moving)  and a range of twenty-seven registration and mesh generation parameters (indicated in Table  of Figure \ref{fig:params}. APBNRR augments PBNRR to accommodate soft-tissue deformation caused by tumor resection. This iterative method adaptively modifies a heterogeneous finite element model to optimize non-rigid registration in the presence of tissue resection. Using the segmented tumor and the registration error at each iteration, APBNRR gradually excludes the resection volume from the model. During each iteration, registration is performed, the registration error is estimated, the mesh is deformed to a predicted resection volume, and the brain model (minus the predicted resection volume) is re-tessellated. Re-tessellation is required to ensure high-quality mesh elements, which is important for the convergence of the linear solver. 

\section{High-Performance Infrastructure for Non-Rigid Registration}

The baseline code used in the design of the NRR  was the implementation
developed and evaluated by Clatz et al.~\cite{clatz05}. Based on the benchmarking and
analysis of this implementation, we identified the following problems:

\begin{enumerate}

\item  The execution time of the original non-rigid registration code is
highly data-dependent. When computed on a high-end 4 CPU workstation, the computation time
varies between 30 and 50 minutes.  The scalability of the code is poor due to
workload imbalances.

\item  The code is designed as a single monolithic component (since it was not
evaluated in the intraoperative mode), and a single failure at any point
requires restarting the registration from the beginning.

\item The original code is implemented  in PVM~\cite{RefWorks:338} which is not widely supported 
as compared to the use of MPI~\cite{RefWorks:337} for message passing.

\end{enumerate}

Consequently, we identified the following implementation objectives in the
design of the system.

 {\bf  High-performance}   Develop  an  efficient  and  portable software
 environment for  parallel and distributed implementation of real-time
 non-rigid  registration   method  for  both small-scale parallel machines
 and large-scale geographically distributed Clusters of Workstations (CoWs).
 The implementation should be able to work on both dedicated and time-shared
 resources.

{\bf Quality-of-service (QoS)} Provide functionality not only to
sustain  failure  but  also  to dynamically  replace/reallocate  faulty
resources  with new  ones during  the real-time  data  acquisition and
computation.

 {\bf Ease-of-use} Develop  a GUI that automatically will handle
exceptions (e.g., faults, resource management, and network outages), and
assist in the parameter initialization.

Different strategies can be explored in the high-performance implementation of the
described NRR method. We first explore how this can be done using ubiquitous
CoWs. During the studies of NRR at BWH, the implementation based on CoW was utilized prospectively, as mentioned in the reference~\cite{archip07}.  We also
describe our  efforts to further increase the availability of the
implementation by developing its components ported on Graphical Processing
Units (GPUs) and studying the use of Grid resources.

We develop NRR   DDDAS based on  the concept  of the 
{\emph{computational    workflow}}.  We re-design the core   NRR
implementation as a  coordinated set of processing components communicating by passing data.  Such an approach allows   to separate
time-critical steps, and concentrate on the optimum parallelization
strategies for   each step  that    requires performance
improvement.

\subsection{Cluster of Workstations}

\begin{figure}[t]
\centering
\includegraphics[width=4.5in,height=4.3in]{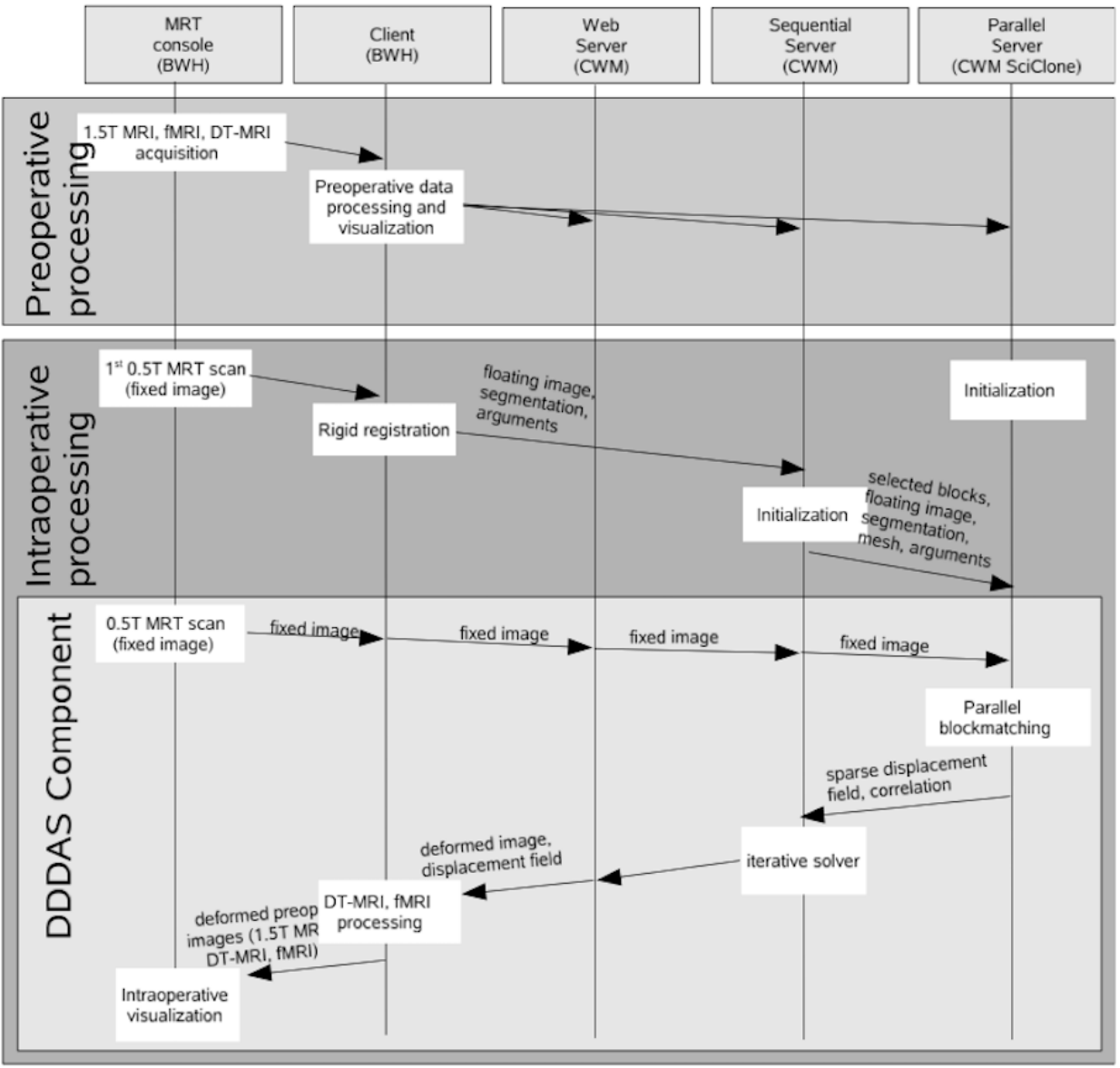}
\caption{\label{fig:timeline} Timeline  of the image  processing steps
during  IGNS (the client is  running at BWH, and the server is using multiple
clusters at CWM, for fault-tolerance purposes).}
\end{figure}

CoWs have become power plants of ubiquitous
computing.  Availability of such cluster  at the College of William and Mary
(CWM, Williamsburg, VA) motivated the development of the implementation of
the  CoW-based NRR  DDDAS.  In addition to the dedicated computing
cluster,  we use the shared resources of a computer lab to boost
the computing power and reliability of the implemented system.  The  targeted
users of our   DDDAS  are clinical researchers of  Brigham and  Women's
Hospital (BWH, Boston, MA). Our approach is to map the components of the
workflow on the computing and communication resources of CWM  and BWH and
expose the DDDAS   to clinical researchers by means of a  web
service interface. The timeline of the interaction with the complete NRR DDDAS
is shown in  Figure~\ref{fig:timeline}. The key component of this system,
which requires parallelization, is block matching.

{\bf {Parallel feature selection.}} 
PBNR and APBNRR algorithmically identify image features by analyzing voxel intensity variation across the intracranial cavity. For each feature candidate, it computes the variance within a block of size  $\mathbf{B_s}$. It then selects $\mathbf{F_s}$   features with the highest variance. Experimental evaluation has shown that when $\mathbf{B_s = 3}$ or 5 and $ \mathbf{ 3 \% < F_s < 10\%}$, a sufficient number of image blocks ($>3×10^5$) can be selected. The method also uses a connectivity pattern to avoid selecting blocks that are too close to each other, thereby influencing the distribution of selected blocks in the image. Three simplex patterns are available: “vertex” (i.e., zero-order simplex implies 26 connectivity), “edge” (i.e., first-order simplex implies 18 connectivity), and “face” (i.e., second-order simplex implies 6 connectivity). The higher the order of the simplex pattern the higher the density of the selected blocks. Since the “face” pattern results in a higher density of blocks near the boundaries/interfaces of anatomical structures, features expected to be most persistent between pre-operative and intraoperative image acquisitions, it is most suitable for IGNS-based Glioma resection. The parallel implementation partitions the preoperative image into k sub-regions, where k is the number of threads. Each thread computes a variance value and an image index for each feature inside the sub-region. Then, the computed pairs are sorted in parallel based on their variance and merged into a global vector. The size of the global vector is equal to the total number of computed blocks. Finally, 0.5 + NFeatures × Fs blocks are selected from the global vector.

{\bf{Multi-level distributed block matching}} To find a match for a given block, we need the block center coordinates and the areas of the fixed and floating images bounded by the block matching window~\cite{clatz05}.   The fixed and floating images are loaded on each processor during the initialization
step, as shown in  Fig.~\ref{fig:timeline}.  The total workload is maintained in a {\emph{work-pool}} data structure. Each item  of the work pool contains the three coordinates of the block center (the total number of blocks for a typical dataset is around 100,000),  and the best match found for that block  (in case the block was processed; otherwise, that field is empty). 

However, because of the scarce resource availability, we have to handle computational clusters that belong to different administrative domains.  We
address this issue with hierarchical multi-level organization of the computation using the master-worker model.   A dedicated master node is
selected within each cluster. The master maintains a  replica of the global work pool and is responsible for distributing the work according to the
requests of the nodes within the assigned cluster and communicating the execution progress to the other master(s).

{\bf{Parallel Mesh Generation}}
The image segmentation generates a patient-specific finite element mesh for PBNRR and APBNRR. The tetrahedral mesh's quality influences the solution's numerical accuracy and the correctness of the estimated transformation. The higher the quality of the elements (i.e., the larger the minimum dihedral angle), the better the convergence of the linear solver. A parallel Delaunay meshing method is employed to tessellate the segmented brain with high-quality tetrahedral elements and to model the brain surface with geometric and topological guarantees \cite{Foteinos14H}. Single-tissue (i.e., brain parenchyma) and multi-tissue (i.e., brain parenchyma and tumor) meshes are generated. Parameter $\delta$  determines the size of the mesh, where a smaller $\delta > 0$ generates a larger mesh.

{\bf{Multi-level Dynamic Load Balancing}} The imbalance of the processing
time across different nodes involved in the computation is caused by our
inability or difficulty to predict the processing time required per block of
data on a given architecture.  The main sources of load imbalance are
{\emph{platform-dependent}}.  These are caused by the heterogeneous nature of
the PEs. More importantly, some of the resources may be time-shared by
multiple users and applications, which affects the processing time unpredictably. The (weighted-) static work assignment is
ineffective when some resources operate in the time-shared mode.

We have implemented a multi-level hierarchical dynamic load balancing scheme
for parallel block matching.  We use an initial rough estimation of the
combined computational power of each cluster involved in the
computation (based on CPU clock speed) for the weighted partitioning of the
work pool and initial work assignment.   However, this is a rough
``guess'' estimation,  which  is adjusted  at  runtime using  a combination of
master/worker and work-stealing~\cite{bib:Blumofe_cilk,bib:Wu_multilist}
methods.  Each master maintains an instance of the global work pool. Initially, 
all these pools are identical. The portion of  the work-pool
assigned  to  a specific cluster  is partitioned into meta-blocks (a  sequence
of blocks),  which are passed to  the cluster nodes using the master-worker
model. As soon as all the matches for a meta-block are computed, they are
communicated back to the master, and a new meta-block is requested.  If
the portion of the work pool assigned to a  master is processed,  the
master continues with the ``remote'' portions of work  (i.e., those
initially assigned to other clusters). As  soon as  the processing of  a
``remote''  meta-block is complete, it is communicated to all the other
master nodes to prevent duplicated computation.

{\bf{Multi-Level Fault Tolerance}} Our implementation is completely
decoupled,  which provides the first level of fault tolerance, i.e., if
the failure takes place at any of the stages, we can seamlessly restart
just the failed phase of the algorithm and recover the computation.  The second level of fault tolerance pertains to the parallel block-matching phase. It is well-known that the vulnerability of parallel computations to
hardware failures increases as we scale the size of the system. We would
like to have a robust system that, in case of failure, could
continue the parallel block matching without recomputing results obtained
before the failure.  This functionality is greatly facilitated by
maintaining the previously described work-pool data structure which is
managed by the master nodes.

The work-pool data structure is replicated on the separate
file systems of these clusters and has a tuple for each block center. A tuple can be either empty if the corresponding block has
not been processed or it contains the three components of
the best match for a  given block.  The work pool is synchronized
periodically between the two clusters, and within each cluster, it is
updated by the  PEs involved. As long as one of
the clusters involved in the computation remains operational, we
can sustain the failure of the other computational side
and deliver the registration result.

{\bf{Ease-of-use}} The implementation comprises the client and server
components.  The  client is running at the hospital site and is based on a
Web service, which makes it highly portable and easy to deploy.  The input data and arguments are transferred to the participating sites on the server side.
Currently, we have a single server responsible for this task. The computation
uses the participating remote sites to provide the
necessary performance and fault tolerance.

\begin{table*}[ht]
\centering
\caption{\label{tab:progress} Response time (sec) of the intra-surgery part
of the CoW-based NRR DDDAS at various stages of development. }
\smallskip
\footnotesize
\begin{tabular}{lccccccc}
\hline\hline
Setup           &\multicolumn{7}{c}{ID}    \\\cline{2-8}
            & 1 & 2 & 3 & 4 & 5 & 6 & 7 \\ \hline
High-end workstation, using original  &1558 & 1850& 2090& 2882& 2317&2302& 3130 \\
PVM implementation      &     &     &     &     &     &    &  \\ \hline
SciClone (240 procs),   &745    &639    &595    &617    &570    &550.4&1153 \\
no load-balancing   & &  & &  &  &  &   \\ \hline
SciClone (240 procs) and    &   &   &   &   &   &   &   \\
CS lab(29 procs), dynamic 2-level&{\bf{30}} &{\bf{40}}  &{\bf{42}}  &{\bf{37}}  &{\bf{34}}  &{\bf{33}}  &{\bf{35}}  \\
load-balancing  and fault-tolerance &   &   &   &   &  &  & \\
\hline\hline
\end{tabular}

\end{table*}

We applied the developed NRR DDDAS for registering seven image datasets
acquired at BWH.  The computations for two of these seven registration
computations were accomplished during the course of surgery (at the  College
of William and   Mary),  while the rest of the computations were
done retrospectively.  All  of the  intra-operative computations utilized
{\emph{SciClone}} (a heterogeneous cluster of workstations located at CWM,
reserved in advance for the registration computation) and the workstations
of the student lab (time-shared mode).  The details of the hardware
configuration can be found in~\cite{sc06nikos}.  Data transfer between the
networks of  CWM and BWH (subnet of Harvard University) is facilitated by
the Internet2 backbone network, with  the slowest link having a bandwidth  of 2.5
Gbps.

The evaluation results are summarized in Table~\ref{tab:progress}.
We were able to reduce the total response time to 2 minutes
(4 minutes, including the time to transfer the data). We showed that dynamic
load balancing is highly effective in the time-shared environment. The modular
structure of the implemented code greatly assisted in the overall usability
and reliability of the code. The fault-tolerance mechanisms implemented
are essential and introduce a  mere 5-10\% increase in the execution
time.

\begin{table}[htb]
\caption{Performance results for the 6 clinical cases with 1 and 12 threads. The experiments were conducted in a workstation with 2 sockets of 6 Intel Xeon X5690@3.47 GHz CPU  cores each, totaling 12 cores and 96GB of RAM. The I/O time is included.}
{\resizebox{\textwidth}{!}{
\begin{tabular}{c c c c c c c c c c c c c c c c}\hline
 & \multicolumn{10}{c}{Time (sec)} & \multicolumn{5}{c}{Speed-up}\\ \hline 
\multirow{2}{*}{Case} & \multicolumn{2}{c}{RIGID} & \multicolumn{2}{c}{BSPLINE} & \multicolumn{2}{c}{PBNRR} & \multicolumn{2}{c}{APBNRR}  & \multicolumn{2}{c}{PAPBNRR} & 
\multirow{2}{*}{RIGID} & \multirow{2}{*}{BSPLINE} & \multirow{2}{*}{PBNRR} & \multirow{2}{*}{APBNRR} & \multirow{2}{*}{PAPBNRR} \\
& 1T & 12T & 1T & 12T & 1T & 12T & 1T & 12T & 1T & 12T & & & & &\\\hline
1&60.18	&16.20	&156.28	&20.08	&138.71	&81.18 &579.62	&213.97			&483.63 &93.14	&3.71	&7.78	&1.71&2.71	&5.19\\
2&280.55&45.31	&128.68	&17.80	&123.76	&73.99 &509.89	&192.70			&275.26	&54.12	&6.19	&7.23	&1.67&2.65	&5.09\\
3&555.65&77.33	&135.91	&18.71	&112.79	&68.72 &486.95	&185.86			&265.61	&52.46	&7.19	&7.26	&1.64&2.62	&5.06\\
4&45.51	&9.76	&33.75	&6.31	&23.80	&20.90&107.43	&63.17			&53.97	&18.51	&4.66	&5.35	&1.14&1.70	&2.92\\
5&52.25	&10.76	&31.26	&5.97	&23.91	&20.62&109.17	&64.08			&53.94	&18.85	&4.86	&5.24	&1.16&1.70	&2.86\\
6&44.53	&8.40	&29.36	&4.61	&19.46	&16.80&81.90	&48.85			&61.32	&21.10	&5.30	&6.37	&1.16&1.68	&2.91\\\hline
Average &173.11	&27.96	&85.87	&12.25	&73.74 & 47.04	&312.49	& 128.11 & 198.96 &43.03 &5.32	&6.54	&1.41 &2.18 &4.00\\
\hline
\end{tabular}}}
\label{performance}
\end{table}

Almost 10 years later, we performed a more comprehensive experimental evaluation at Old Dominion University using a Dell Linux workstation with two sockets of six Intel Xeon X5690@3.47 GHz CPU  cores each, totaling twelve cores and 96GB of RAM. For the Rigid and BSpline, we run the BRAINSFit module from the terminal otherwise, Slicer's GUI degrades the performance significantly. Table~\ref{performance} from~ \cite{Drakopoulos15A} illustrates the end-to-end execution time (including I/O) and the speed-up.   Slicer's BRAINSFit module exhibits a real-time performance with twelve threads.  The BSpline is the fastest among all the methods requiring on average 12.25 seconds to complete. Also, it is highly parallelizable, with an average speed-up of 6.54.  The corresponding values for the Rigid registration are 27.96 seconds and a speed-up of 5.32.

The Parallel APBNRR  (in short PAPBNRR) refers to the additional parallelization of APBNRR modules added and presented in~\cite{Drakopoulos15A} and bring the performance of the adaptive physics-based method to nearly real-time.  Nevertheless, our latest non-rigid registration technology presented in detail in \cite{Drakopoulos15A}  provides the image alignments extremely fast (between 18.51s and 93.14s) for two reasons: (i) it requires fewer adaptive iterations to register the MRI volumes compared to the APBNRR method,  and (ii) it exploits additional parallelism. Indeed, the PAPBNRR completes the registration on average $47.04/43.03\approx 1.1$ and $128.11/42.60\approx 3$ times faster than the PBNRR and APBNRR, respectively. 

According to our study in~\cite{Drakopoulos15A}, the combination of algorithms, software, and hardware has significantly improved real-time results in 10 years since we started this project. In 2015, a single high-performance computing workstation with 6 CPUs at ODU (vs. 270 CPUs used at CWM in 2005-6) was on average 2.93 times more accurate than BSpline, 3.12 times more accurate than PBNRR, and 3.78 times more accurate than Rigid registration.

\subsection{Grid Computing Resources}

Before the widespread use of multi-core CPUs and heterogenous workstations (or nodes)   the HPC community experimented with Grid Computing, which has since evolved into Cloud Computing. This idea gained popularity around 2008, the same year the first draft of this chapter was written.So during that time, the HPC community invested significant effort in developing standards and software for Grid computing, deploying production grid systems worldwide and porting applications on those systems. One such production system under continuous improvement and development is USA-based TeraGrid~\cite{teragrid}. As of May 2007, TeraGrid was connecting 11 high-end computational sites within the USA, providing ``...more than 250 TFLOPS of computing capability and more than 30 petabytes of storage'' and therefore making TeraGrid ...`` the world's largest supercomputer (at that point), most comprehensive distributed cyberinfrastructure for open
scientific research''~\cite{teragrid}.  Around 2008, TeraGrid connected 11 computational centers to provide a cumulative peak performance of 1124 teraflops. 

Using the Grid infrastructure for NRR DDDAS had two major advantages. First, the implementation is not restricted to run on a
specific cluster resource. With the multiple computing centers participating
in TeraGrid, temporary resource outages are more feasible. Second,
complex image processing methods, like NRR, often require the proper setting of
many parameters to achieve optimum accuracy.
Identification of such parameter combinations is a non-trivial task.
One approach to selecting the optimum parameter combination is to use
{\emph{speculative computation}}~\cite{RefWorks:927}, when multiple instances of NRR are
computed in parallel with different parameter settings. In this regard,
we have developed initial accuracy assessment solutions~\cite{RefWorks:1003} to facilitate
intra-operative speculative NRR over the Grid. In out Grid NRR DDDAS, we
leverage the CoW-based implementation, augment it with the automatic error
estimation, and develop a framework for the speculative execution of NRR on the
TeraGrid.

While TeraGrid resources could be accessed directly for individual job
submission and data transfer, doing this manually on a large scale or as part of
workflow execution is not practical.  We adopted Swift workflow scripting and
management system~\cite{RefWorks:984} to implement and deploy NRR workflow.
Swift has been developed and evaluated to support grid implementations based on Globus Toolkit, which allows to use this system without any
modifications to schedule workflows on TeraGrid.  SwiftScript, the scripting
language used for workflow definition, is a powerful way of abstracting
interaction of the processing tasks, which allows defining composite data
inputs, and dependencies between the processing tasks and provides familiar
control structures like loops and conditional structures, which allow flexible
control over workflow definition and execution.

Fault tolerance and dynamic load balancing are important characteristics of
NRR DDDAS. Swift implements basic fault tolerance of workflow execution at the
individual task level, which is critical for NRR computations. If a
particular task fails to deliver the output, Swift will re-schedule its
execution, possibly on a different site.  The Swift infrastructure also provides task-level load-balancing.  The execution traces for the same
computational task are continuously collected and used to dynamically select
the best-performing site when the task is scheduled again.

Swift provides the means to define and execute the workflow, which consists of
individual processing tasks. Each of the processing tasks must be available as
an executable at each of the sites that will be involved in the workflow
computation. The details of running a specific task are provided to Swift in
the so-called {\emph{translation catalog}} available at the client
(submission) site.  The translation catalog contains the identifier of the
remote site where the executable is installed and the optional
information on its invocation. 

\begin{figure}[h]
\includegraphics[width=5in,height=2.0in]{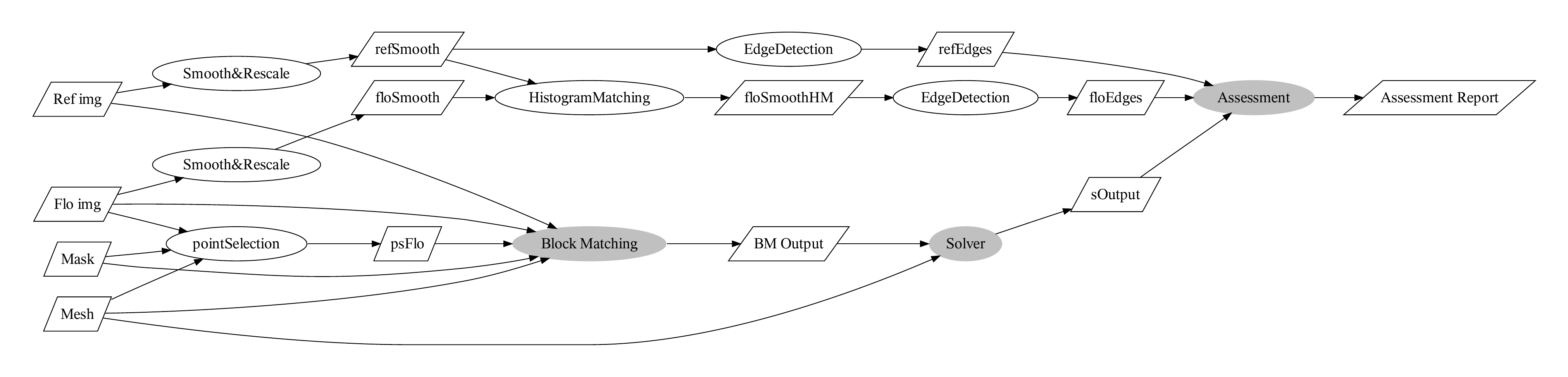}
\caption{NRR workflow diagram for single registration execution (shaded are
the time-critical components of the workflow).\label{fig:nrr_single}}
\end{figure}

The NRR workflow diagram of a single NRR procedure, together with the accuracy
assessment module, is shown in Figure~\ref{fig:nrr_single}.  The block-matching
task is parallelized using MPI and deployed on the TeraGrid sites
for remote parallel execution. The other workflow components are
executed on the local resources (single node of the CWM SciClone cluster).
This NRR workflow corresponds to the base case for computation supported by
the cluster-based implementation we discussed earlier, augmented with the
accuracy assessment module. The accuracy assessment module of the Grid-based
NRR DDDAS was developed separately~\cite{RefWorks:1003}. This module allows to estimate the registration error automatically.  The construction of the
workflow for speculative execution is straightforward with the scripting
capabilities of Swift. This allows us to study some parameters' impact on registration accuracy.

\begin{table}
\caption{Absolute improvement in accuracy (mm) evaluated at selected landmarks
using optimal values of block size and outlier rejection rate.}
\begin{tabularx}{\linewidth}{cXXXXXXXXXXXX}\hline\hline
Case  & 1    & 2    & 3    & 4    & 5    & 6    & 7    & 8    & 9    & 10   & 11  & 12  \\\hline
1     & 0.2  & 0.1  & 0.2  & 0.2  & 0.3  & 0.1  & 0.3  & 0.4  & 0.0  & 0.2  & 0.2 & 0.3 \\
2     & 0.2  & 0.0  & 0.0  & 0.0  & 0.2  & 0.3  & 0.3  & 0.2  & 0.5  & 0.3  & 0.1 & 0.2 \\
3     & 0.6  & 1.0  & 0.2  & 2.9  & 0.0  & 0.1  & 0.3  & --  & --  & --  & -- & -- \\
4     & 0.9  & 0.8  & 0.2  & 0.4  & 0.7  & 0.5  & 0.7  & 0.4  & 0.3  & 0.7  & 0.7 & -- \\
5     & 0.0  & 0.3  & 0.0  & 0.0  & 0.8  & 0.5  & 0.4  & 0.0  & 0.3  & 0.4  & -- & -- \\
6     & 0.2  & 0.1  & 0.2  & 0.1  & 2.0  & 0.1  & 0.0  & 0.1  & --  & --  & -- & -- \\\hline\hline
\end{tabularx}
\label{tab:lm_improvement}
\end{table}

\begin{figure}
\centering
\begin{minipage}{\textwidth}
\leftfigure{\includegraphics[width=.5\linewidth]{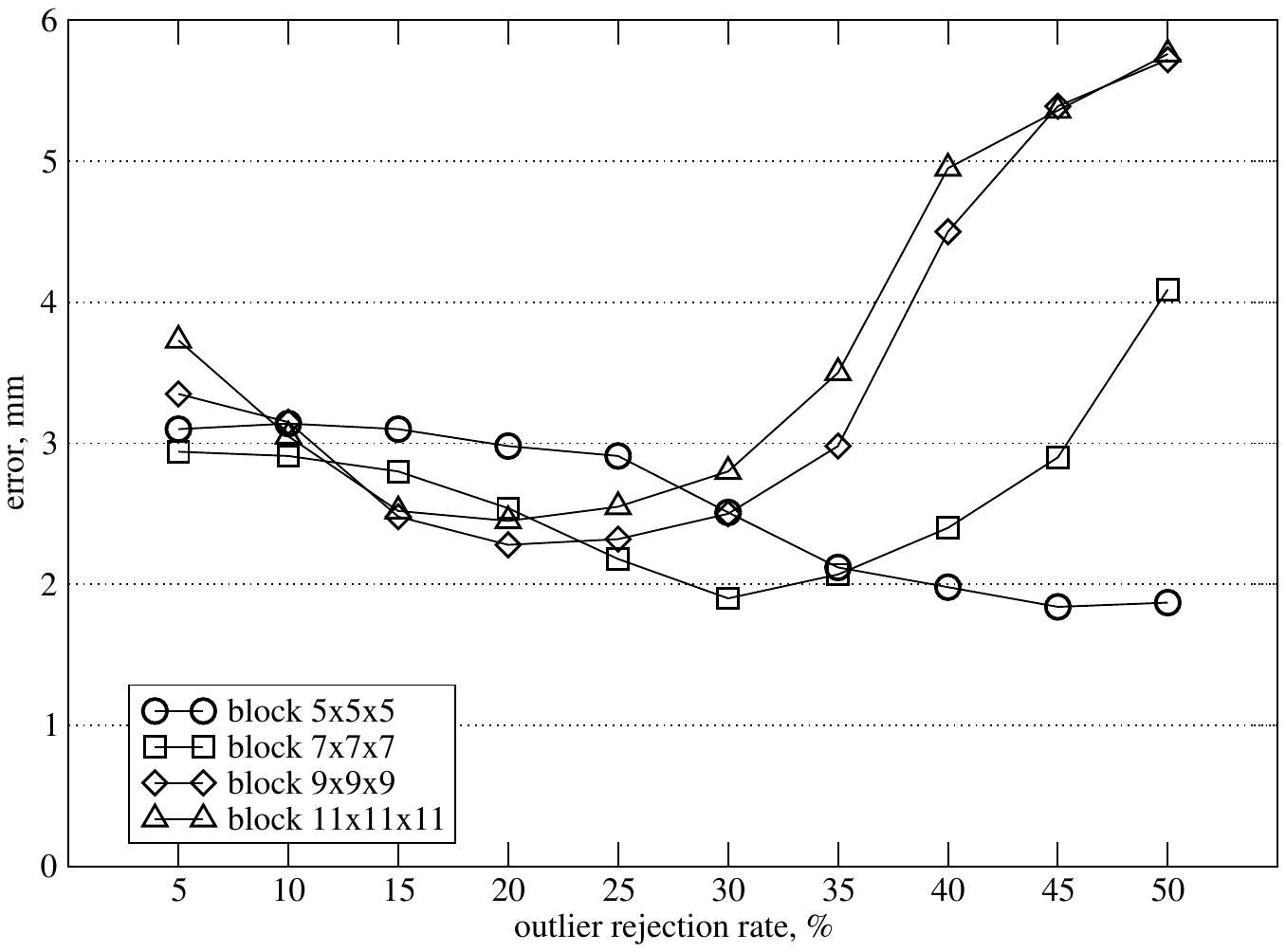}}\hspace{\fill}
\rightfigure{\includegraphics[width=.5\linewidth]{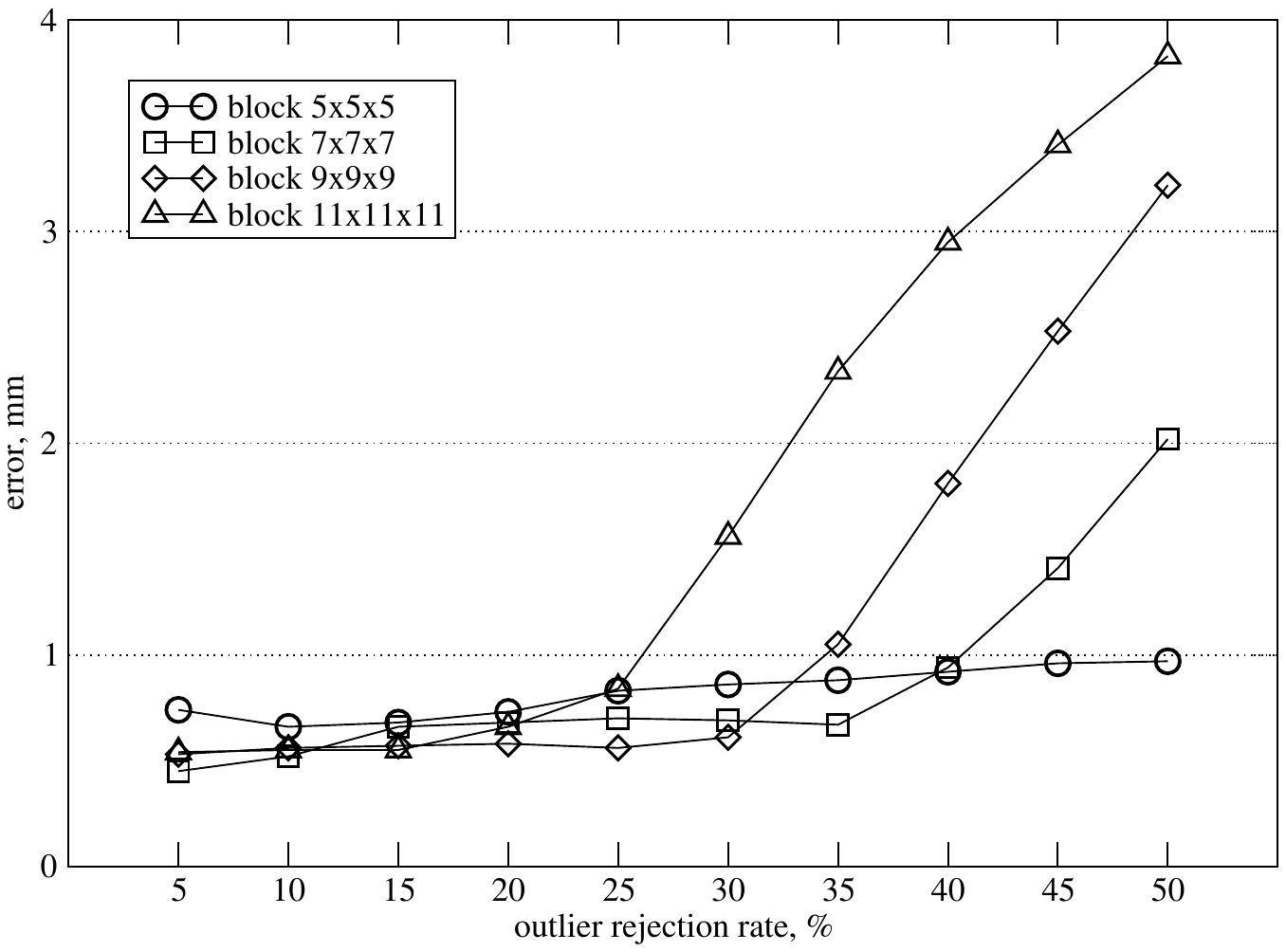}}
\end{minipage}
\caption{Influence of the block size and rejection rate on landmark error:
case 1, landmark 5 (left), and case 1, landmark 12 (right).}
\label{fig:lm_intra}
\end{figure}

We considered the impact of varying the block size and outlier rejection rate
on the accuracy of NRR on retrospective clinical data.
Table~\ref{tab:lm_improvement} summarizes the improvement in accuracy
evaluated at the expert-selected anatomical landmarks with the optimum
combination of these two parameters compared to their default settings.
Based on the experimental data, in most cases, good registration accuracy is
achieved using the default parameters suggested by Clatz et
al.~\cite{clatz05}.  However, in Case 3, the improvement in registration
accuracy was significant. In both cases, however, there were landmark points
where registration errors exceeded voxel dimensions.  The analyzed data also
suggests that the optimum value of outlier rejection is varied in different
locations of the image.  For example, if we consider landmarks 5 and 12 in
Case 1, the optimal combination of the studied parameters differs in each
case, as shown in Figure~\ref{fig:lm_intra}.

\subsection{Graphical Processing Units}

Graphics Processing Unit (GPU), an inexpensive, single-chip, massively
parallel architecture, has shown higher throughput and
performance per dollar orders than traditional CPUs. In addition, a GPU can be easily
deployed in the Operating Room as a co-processor of the CPU without hindering
routine surgical operations.  Around 2008, researchers have made
efforts to accelerate NRR using 
GPU~\cite{AntGPUNRR08, MuyGPUNRR08,VetGPUNRR07, LevGPUNRR04}.
However, to satisfy the requirement for the
accuracy and real-time in the clinic, a more advanced GPU-based NRR is imperative.

The workflow implementation of NRR DDDAS allows us to use the best parallelization
strategies for individual components. The block matching component is
embarrassingly parallel, which makes it highly amenable to GPU parallelization.  
We use CUDA programming model~\cite{cuda} to develop the GPU implementation of
this component. CUDA organizes GPU threads in the grid, which is an array of blocks
and each block is an array of threads. The kernel is the core code to be
executed on each thread, which performs on different data sets using its ID
in a SIMD fashion. CUDA programming model can be treated as two levels loop:
block level and thread level. In the following code, the outer loop can be
parallelized using GPU on the thread block level. Computation of the
similarity metric for an image block (inner loop) is parallelized on 
GPU thread level, while the similarity metric computation is done on CPU:

\begin{algorithmic}[1]
\FOR{each image block $fblk$ in floating image}
\STATE define search window $sw$ in fixed image
\FOR{each image block $tblk$ in $sw$}
\STATE calculate similarity $s$ between $fblk$ and $tblk$
\ENDFOR
\STATE find the maximum $s$ and corresponding displacement
\ENDFOR
\end{algorithmic}

\begin{figure}[bt]
\centering
\begin{minipage}{\textwidth}
\leftfigure{\includegraphics[width=.5\linewidth]{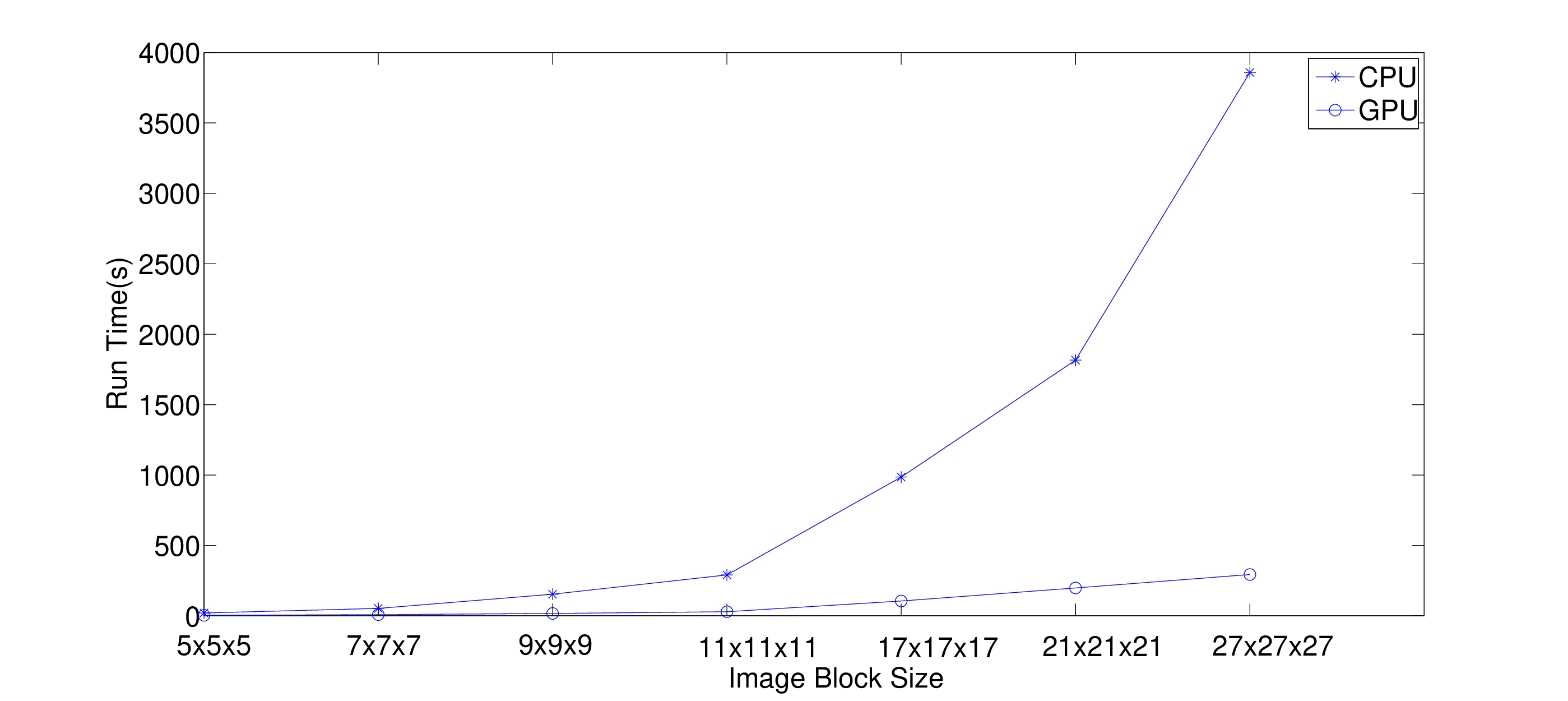}} \hspace{\fill}
\rightfigure{\includegraphics[width=.5\linewidth]{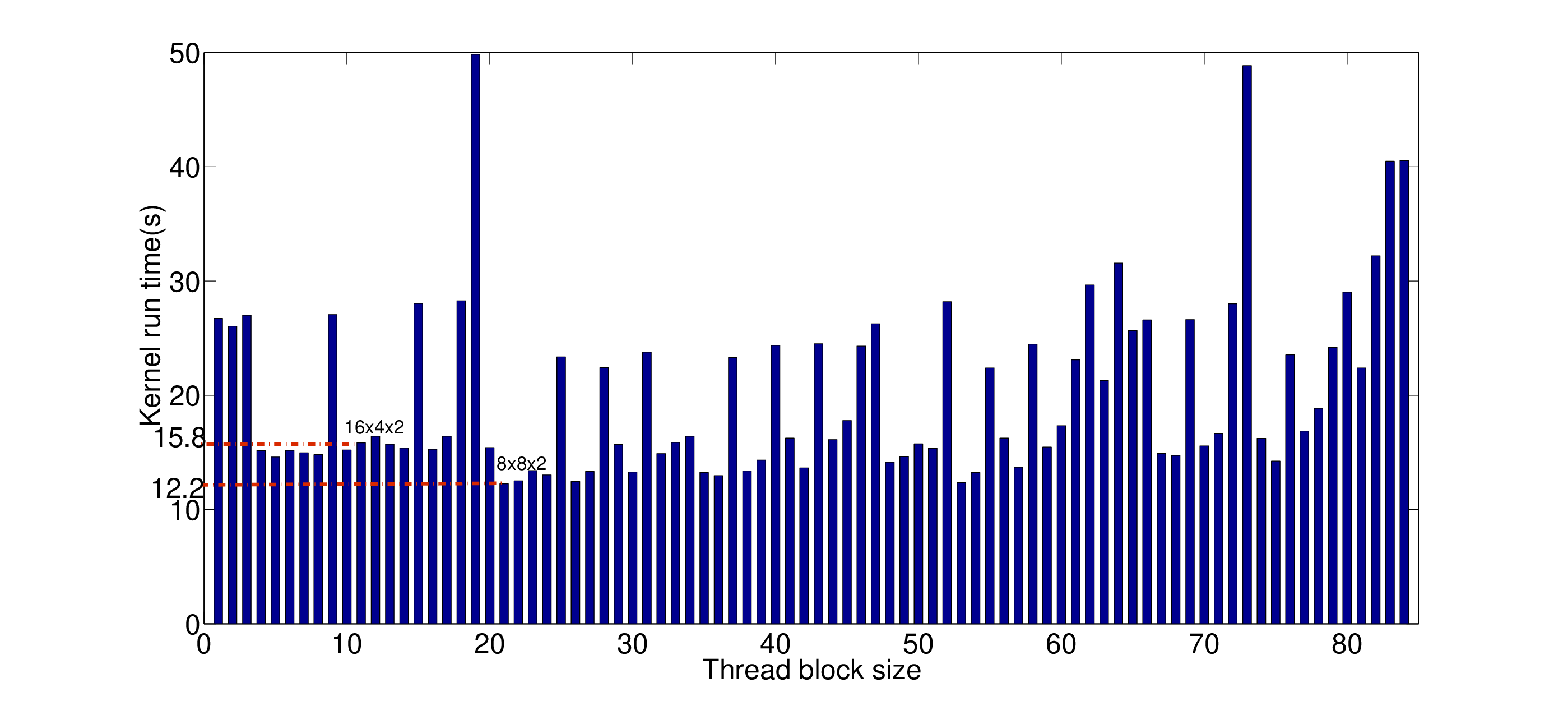}}
\end{minipage}
\caption{Parallel block matching execution time on CPU vs. GPU (Left) and the effect of GPU thread configuration on the performance of the block matching
kernel code (Right). }
\label{fig:BMGPUCPU}
\end{figure}

This GPU-based implementation of block matching can gain a speedup of about
10, as we show in Figure~\ref{fig:BMGPUCPU}, compared to CPU.  The speed-up is
measured at seven different image block sizes. Figure~\ref{fig:BMGPUCPU}
clearly shows that GPU running time increases linearly as we increase the
block size, but CPU exhibits a super linear behavior.


Optimization of GPU codes is particularly important since there are numerous
parameters of the execution environment, which can affect performance.
Significant evidence exists that there can be orders of magnitude performance
differences depending on the level of optimization for GPU implementations
~\cite{GPUOptICS,GPUOptPPoPP,GPUOptCGO}.  The search space generated by the
execution configuration is so large that finding the
optimal parameters by trial and error is not practical.  Several studies (around 2008)  tackled this
problem through empirical search-based approaches~\cite{GPUOptCGO,GADAPT}. We
utilize the method provided in~\cite{GADAPT} to optimize the GPU execution
configuration for block matching and improve speedup further, as shown in
Figure~\ref{fig:BMGPUCPU} (right).  We observe a speedup of about 30 when comparing
optimized and non-optimized implementations.


\section{Deep Learning Model Architecture}

Through the evolution of Grid Computing to Cloud computing and advances in AI, we have made significant progress in improving the accuracy and execution time of NRR by leveraging Machine Learning for Patient-Specific search of NRR parameters. In this section,  we summarize our work appeared in \cite{Angelopoulos19D}, where we have presented a model that uses a deep feedforward neural network trained on data from past APBNRR executions. {\em The model takes fourteen parameters as input, including twelve APBNRR parameters with a very large range of possible values and two additional patient-specific parameters.} The output is a predicted Hausdorff distance of the registered preoperative image. The idea is to replace speculative executions over the Grid (now Cloud) we described earlier in Section 4.2. The twelve APBNRR parameters are the half block size $\mathbf{B_s}$, the half window size $\mathbf{W_s}$ in the axial, coronal, and sagittal direction, the fraction $\mathbf{F_s}$ to select the image blocks, the number of approximation (outlier rejection) steps $\mathbf{N_{rej}}$, the number of interpolation steps $\mathbf{N_{int}}$, the maximum number of adaptive iterations $\mathbf{N_{(iter,.max)}}$, the minimum number of blocks with a zero correspondence $\mathbf{N_{(b0,min)}}$, and the percentage of rejected outlier blocks $\mathbf{F_r}.$ The two patient-specific parameters are the location of the tumor in the brain (lobe-wise) and the degree of brain deformation, which can be directly inferred from the rigid registration error. These two parameters improve model performance by providing additional cues for the neural network to learn and fine-tune the model for a specific patient during an IGNS session. 

The neural network was implemented using Keras \cite{chollet2015keras} on a TensorFlow backend \cite{abadi2016tensorflow}. It consists of four hidden, fully connected layers, each comprised of 128 neurons. We used ReLU \cite{arora2018understanding} as the activation function and stochastic gradient descent with Nesterov momentum \cite{ruder2017overview} for optimization. We chose to use SGD because it has been shown to lead to better generalization compared to adaptive gradient optimizers such as Adam, due to its tendency to converge to better global minima \cite{keskar2017improving}. No dropout layers or methods of preventing overfitting were used, as the training data is complex and highly variated. Furthermore, the feature selection in parallel mode is non-deterministic, meaning that the same input parameters can yield different outputs in every execution, 
further reducing the ability of the model to “memorize” the training data and overfit to the training set. The neural network was built in an iterative fashion, wherein at each iteration, a new patient case was added (consisting of data from about ~100,000 APBNRR executions) and the network’s hyperparameters and architecture were tuned to reduce the loss marked in the previous iteration. 

\begin{figure}[h]
\includegraphics[width=.99\linewidth]{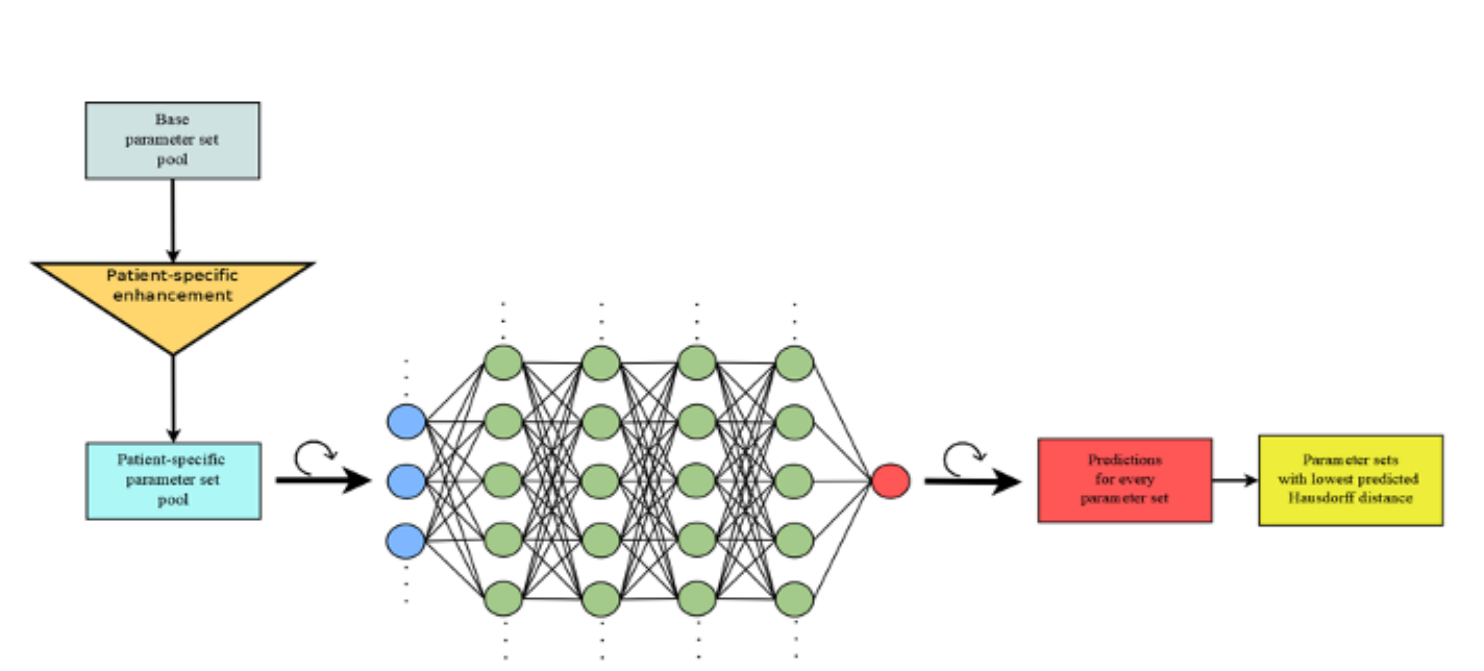}
\caption{Deep learning portion of APBNRR registration. \label{fig:deep-learning}}
\end{figure}

\subsection{ Deep Learning APBNRR}

The deep learning portion of APBNRR takes place before the actual execution of APBNRR. APBNRR inputs the parameter sets predicted by the deep learning model to result in the lowest Hausdorff distances. A visualization of how the deep learning model works is shown in  Figure \ref{fig:deep-learning}. The neural network is given as input each parameter set in a pool consisting of patient-specific parameter sets, which was produced by augmenting a base, general parameter set pool. We have created tools that do this automatically. The neural network iterates through each parameter set and outputs the Hausdorff distance of the registered image that APBNRR would produce if this parameter set were utilized. Out of all those predictions, the lowest ones are compiled in a file and can be used as input to APBNRR. This process takes about 15 seconds.

The deep learning model takes input parameter sets from a pool specific to each patient. This is an augmented version of a base, general parameter pool, which has been enhanced to include two patient-specific features: the tumor's location in the brain and the size of the deformation, as can be derived from the rigid registration error. In our experiments, the patient-specific features have yielded significantly better results.

We encountered a couple of important challenges while constructing the deep learning model. The first is that APBNRR is non-deterministic due to the parallel feature selection algorithm. Given two identical parameter sets, APBNRR will yield two registered images with different Hausdorff distances. This non-determinism leads to a degree of “randomness” in results, which hinders the ability of the neural network to predict the Hausdorff distances correctly. Unfortunately, there is no practical solution to this issue other than amassing more training data to “average out” this randomness, leading us to the second challenge: collecting training data takes a long time. To gather training data, we must run APBNRR for every 
parameter set in an exhaustive pool for every additional patient case we want to include in our training set. Evaluating a single case exhaustively (~1 million executions) takes about 1-2 months, depending on the severity and size of the brain tumor.

\subsection{Final Experimental Evaluation}
This section summarizes our findings in \cite{Angelopoulos19D}  and a summary in \cite{Drakopoulos21A}. Our data set for the model consists of medical resonance images from thirteen patient cases provided by the Huashan Worldwide Medical Center and Brigham and Women’s Hospital of Harvard University and output data from over 2.6 million executions of APBNRR, collected over five months. Eleven of the thirteen cases were used for training (~2.4 million parameter sets) and two for evaluation (~200,000 parameter sets). APBNRR was executed with arrays of 120 parameter sets on a supercomputing cluster, utilizing 450 CPU cores and over 4 TB of RAM. The total computing time required for the executions was approximately 3600 hours. 

\begin{figure}[h]
\includegraphics[width=.99\linewidth]{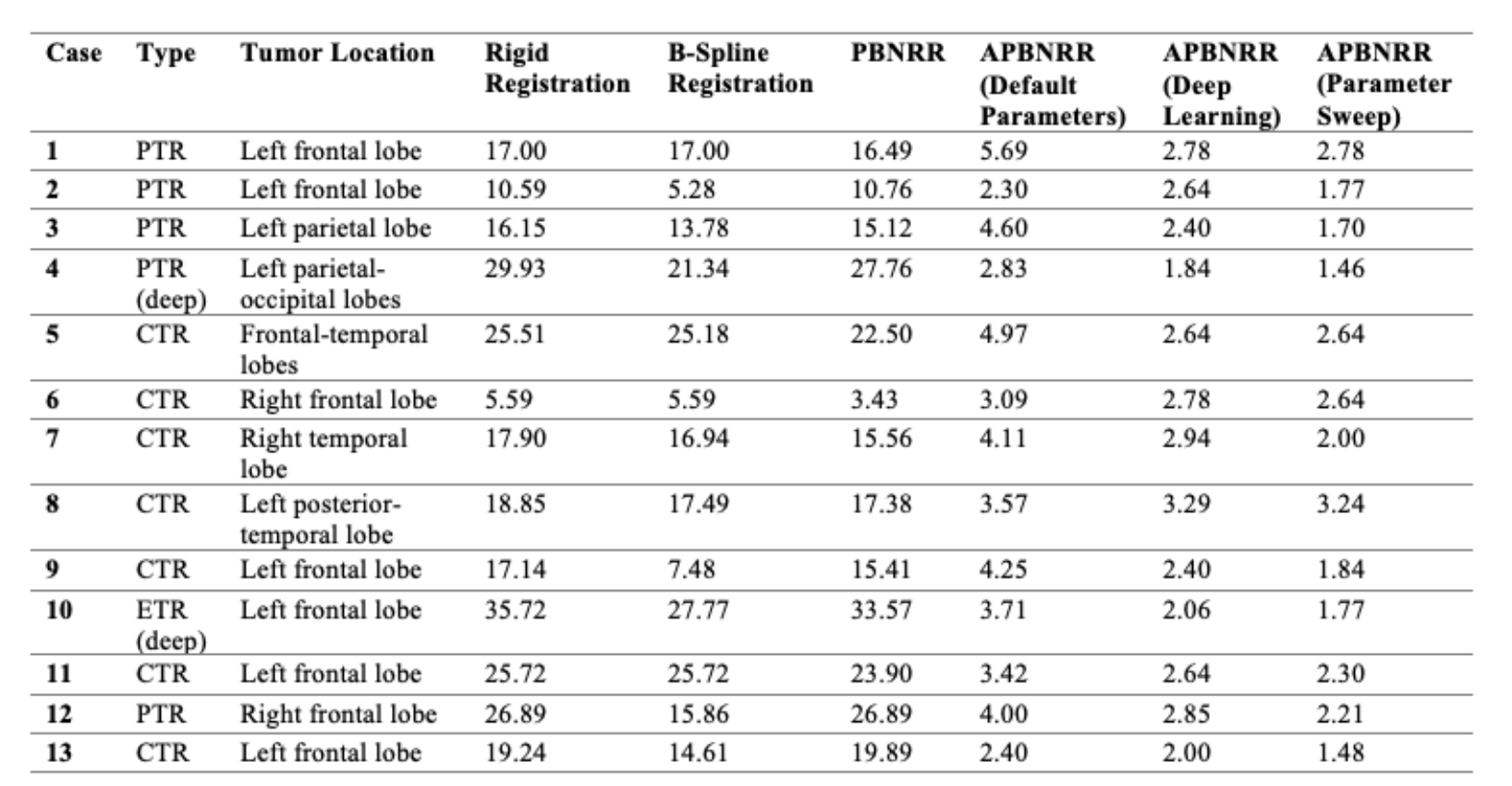}
\caption{Shows the thirteen patient cases that consist of our data set and the results (measured as the Hausdorff distance in mm) achieved with various methods of registration, including with APBNRR using deep learning and APBNRR using a parameter sweep. Cases with numbers 1-11 were used for training, and 12, and 13 were used for evaluation.) \label{fig: results}}
\end{figure}

We used the deep learning model to achieve a training root mean squared error (RMSE) of 1.41 and an evaluation RMSE of 1.21 for predicted Hausdorff distances. For further evaluation, utilizing the trained model, we again executed APBNRR for the 13 cases, using the top 120 parameter sets predicted by the model for each case to yield the lowest Hausdorff distances. The best results of those executions are displayed in Table of Figure \ref{fig: results}, along with the results from several other registration methods as a comparison. The best results from the APBNRR parameter sweeps that were used for generating our dataset are also displayed as a reference of what could have been the lowest Hausdorff distance value. 

On average, APBNRR with deep learning is ~8.45 times better than rigid registration, ~6.71 times better than B-Spline registration, and ~7.9 times better than PBNRR. Overall, APBNRR with deep learning leads to superior results (for tested cases) than any of the registration methods noted above. Choosing the correct parameters for medical image registration is a difficult task, as there are many (usually infinite) possible values and combinations of parameters that can lead to better or worse results. The deep learning portion of the APBNRR framework makes this faster and easier by greatly limiting the set of possible optimal parameters for each patient, moving APBNRR one step closer to being utilized in a real-time setting where registration accuracy and speed are critical.

\section{Discussion}

We have briefly described the use of  Dynamic Data Driven Application Systems for Image-Guided Neurosurgery, enabled by the advances in medical image
acquisition, parallel/distributed computing, Machine Learning, and algorithmic changes in \cite{Drakopoulos14A, Drakopoulos15A, Drakopoulos17A}  the original NRR method presented in \cite{clatz05}.  This DDDAS concept, {\em for the first time in clinical practice \cite{sc06nikos, archip07}, helped us to complete and present non-rigid registration results to neuro-surgeons \cite{archip07} at BWH during tumor resection procedures using image landmark tracking across the entire brain volume.}  This is accomplished in three phases:

\begin{itemize}
\item{\bf Phase I:} Reduced the total response time of the time-critical computation component to about 35 sec, delivering an effective speedup of nearly 100 compared to the original implementation \cite{clatz05}.  To achieve this, we used remotely  (at   CWM)    several    CoWs with a total of  269 processors~\cite{sc06nikos} (the data transfer from CWM to the operating room in BWH takes between  3 to 5 min).  Our data suggest that we improved the accuracy of the NRR method by performing speculative execution~\cite{wm-tr-2009-05} on the TeraGrid, which was capable of delivering about  250  TFLOPS. This computing power translates into tens of thousands of registrations (with different parameters) in almost real-time if there was proper coordination with all sides to avoid scheduling conflicts. However, this requires the availability of a network connection between the operating room and remote computing resources. Also, the imaging data must be anonymized before transfer to address confidentiality concerns. 

\item{\bf Phase II:} Eliminated the use of Grid and lately Cloud computing for the speculative execution of PBNRR by using much cheaper hardware that can be located nearby the operating room.   Our results in the  CRTC indicate that it is possible to complete the time-critical component of non-rigid registration within a minute ---save another  3  to 5  minutes for the data transfer--- using a  single (or two, for fault-tolerance)   high-end workstations with NVIDIA GeForce 8800  GT GPU and   2 x Intel Core2 Duo  CPU 3.16GHz.     We believe in using current and emerging heterogeneous hardware architectures and the coordinated use of  Cloud (or TeraGrid when performing these studies). Our results show that GPU provides excellent computing capabilities without sacrificing the result's accuracy. 

\item{\bf Phase III:} Improved the accuracy of the original NRR method by extending the NRR model to incorporate adaptivity to improve accuracy and replaced the expensive speculative execution of multiple NRRs (from Phase I \& II ) with machine learning to improve both accuracy and performance. Our results indicate that the APBNRR with deep learning is ~8.45 times better than rigid registration, ~6.71 times better than B-Spline registration, and ~7.9 times better than PBNRR (outcome of Phase I and II). A key element for evaluating the performance of APBNRR is determining optimal input parameters. Optimal input parameters are usually in our base parameter pool's first 800 – 39,000 parameter-sets. Determining optimal input parameters by doing a parameter sweep utilizing arrays of 120 jobs over the Cloud would take an estimated 33 minutes – 27 hours.  This high variability is not acceptable in a neurosurgery session. Deep learning offers significant speedups in this area by allowing the user to limit the parameter pool to a custom value. By using deep learning, we could evaluate fewer parameter sets, potentially completing APBNRR in 5 minutes on moderate CoWs (5 to 10 HPC nodes) while also producing good results.

\end{itemize}

This Chapter presented only a few of the most important findings and results. However, several efforts at CRTC alone address other aspects like: (1) 
 robust and real-time image-to-mesh (I2M) conversion \cite{Fedorov05T, Rivara06P, Joshi07A, Joshi08A, Fedorov08A, Chrisochoide08P, Fedorov08T, Foteinos09T, Foteinos10G, Foteinos10A, Liu10M, Audette11A, Chernikov11M, Chernikov11T, Foteinos11H, Foteinos11A, Foteinos12H, Foteinos12D, Foteinos12M, Liu12M, Chernikov13T, Foteinos134, Foteinos13H, Foteinos14G, Foteinos14H, Drakopoulos15T, Foteinos154, Drakopoulos15I, Feng15S, Drakopoulos15T, Drakopoulos16L, Garner23T},  (2) more accurate computation of Hausdorf Distance \cite{Billet08T, Fedorov08E}, (3) parallel Euclidean Distance Transform, which is critical for parallel I2M conversion \cite{Staubs06P}, and many preliminary efforts that lead to the findings presented in this Chapter and we list them here for completeness without  expanding on them \cite{Archip06I, Fedorov06A, Chrisochoides07G, Chrisochoide07R, Liu09R, Liu10N, Weissberger10M, Liu10A, Liu10B, Liu10U, Liu11M, Liu12A, Liu12H, Liu14A, Liu14B, Drakopoulos15A,  Garner15I, Drakopoulos16B, Mohrehkesh16L, Kazakidi16N, Drakopoulos16A, Drakopoulos17A, Best19I, Drakopoulos21A}

\section{Conclusion and Future Work}

Choosing the correct parameters for medical image registration is a difficult task, as there are many (usually infinite) possible values and combinations of parameters that can lead to better or worse results. The deep learning portion of the APBNRR framework makes this easier by greatly limiting the set of possible optimal parameters for each patient, moving APBNRR one step closer to being utilized in a real-time setting where registration accuracy and speed are critical.

Operating rooms, like Advanced Multi-modality Image Guided Operating (AMIGO) ~\cite{amigo} suite, provide new capabilities to improve intra-operative image guidance. Advances in high-performance and distributed tools for image analysis, like the NRR DDDAS we presented in this chapter, will be essential to meet the ever-increasing computational demands of such environments. DDDAS will be critical in health care, among other areas, where this concept proved successful~\cite{dddas, dddas_book}. 

Within AMIGO's workflow, APBNRR execution takes place after the intraoperative MR phase and is meant to assist in tumor and residual tumor assessment due to the ability of APBNRR to perform well in the context of tumor resection. The registered intraoperative image with the tumor resected allows the neurosurgeon to evaluate better how well the tumor has been resected. The time window for APBNRR to execute as part of this workflow is a few minutes. As such, on top of accuracy, speed is also critical. Deep learning solves the problem of limiting the parameter pool that needs to be evaluated, but some issues with APBNRR need to be resolved. One of the most important is the dependence of APBNRR performance on the tumor size. The larger the tumor, the more time it takes for APBNRR to execute. Furthermore, the more adaptive iterations APBNRR goes through, the longer the execution. We have seen an execution time ranging from ~5 to ~30 minutes --for large and deep brain tumors, with parameter sets having more adaptive iterations and slower execution than those with lower ones but better registration accuracy on average. This suggests we still need to evaluate deep learning APBNRR on how it would perform in a real-world setting.  Specifically, we will look at how APBNRR would perform in the brain tumor resection workflow in AMIGO,  which represents an ideal setting for using APBNRR.

Deep learning APBNRR takes us a step closer to enabling NRR to be used in real-world scenarios. However, some issues must be resolved before that can happen. First, more training data must be collected for the deep learning model to offer more accurate predictions. Second, work needs to be done to enable the deep learning model to generate a parameter pool that is limited in size and can also be evaluated rapidly. Finally, the performance of APBNRR needs to be further improved and extensively evaluated for deep tumors which involve very large brain deformation. Finally, the overall performance of deep learning APBNRR depends on the computational resources available. In particular, the more computational resources available, the larger the parameter pool of the best parameter sets predicted by the neural network can be. As a result, there is a greater chance of achieving better accuracy and performance.

\section*{Acknowledgments} 
The Richard T. Cheng Endowment supports NC.  AF, YL, AK, and NC were partly supported by the NSF grants  CCS-0719929,  CCS-0750901, CCF-0833081, CCF-1439079, and the John   Simon Guggenheim  Memorial Foundation. AF was supported in part by a grant from Brain Science Foundation. AF was supported in part by a grant from Brain Science Foundation. RK, AG, and PB were supported in part by NIH grants: U41 RR019703, P01 CA67165, NIH R01NS049251, NIH 5R01EB027134, and NCIGT P41EB015898.  This research was supported by an allocation through the TeraGrid  Advanced  Support Program.  This work was performed   [in part]   using computational facilities at the  College of William and Mary and Old Dominion University, which were enabled by grants from Sun  Microsystems, the National Science Foundation,  and Virginia's Commonwealth Technology Research Fund. We also thank Dr. Frederica Darema for her hard work laying the foundation, building a DDDAS community, and encouraging us to present our work in this Volume. 

\bibliographystyle{plain}
\bibliography{parallel,igns,gpu_ref}

\end{document}